\documentstyle[psfig]{l-aa}
\begin{document}
\thesaurus{08.02.1, 08.02.2, 08.09.2: YY CMi} 
\title{YY CMi : contact or near contact system?
\thanks{Based on observations partly made at the European Southern Observatory
(ESO)}}
\author{P.G. Niarchos\inst{1} \and L. Mantegazza\inst{2} 
\and E. Poretti\inst{2} \and V. Manimanis\inst{1} 
}
\offprints{P.G. Niarchos}
\institute{
Section of Astrophysics, Department of Physics, University of Athens,
Panepistimiopolis, GR-15784 Zografos, Athens, Greece
\and
Osservatorio Astronomico di Brera, Via Bianchi 46, 23807 Merate (LC), Italy}
\date{Received\dots, accepted\dots}
\maketitle

\begin{abstract}
New V photoelectric observations of the eclipsing system YY CMi, obtained
at La Silla, Chile, and Merate Observatory, Italy, are presented. New times 
of minima and ephemeris based on our observations are also given. The V light
curve was analysed by using the WD code to derive the geometrical and 
physical parameters of the system. Since no spectroscopic mass ratio is 
available, the $q$-search method was applied to yield the preliminary range
of the mass ratio in order to search for the final solution. First the 
unspotted solution was carried out by using the unperturbed parts of the light
curve and applying the DC program of the WD code. The solution was performed
by assuming contact (mode 3) and semi-detached (mode 4) configuration, since 
no classification of the system is possible from the shape of the light curve. 
The solution in mode 4 does not lead to an acceptable model, since the 
secondary was found to be slightly overcontact. Therefore the contact solution 
was finally adopted. Moreover the light curve peculiarities ($\rm Max\,II$
fainter 
than $\rm Max\,I$ and excess of light around the phase 0.32) were explained by 
assuming a cool and a hot spot on the surface of the secondary (cooler) 
component. The degree of contact is very small $(f\approx 3\%)$ and the thermal
contact 
is poor $(T_1-T_2)\approx 650 K$. These  results together with the high
photometric mass ratio $q\approx 0.89$ indicate that YY CMi is very probably
a system at the beginning or the end of the contact phase.\\

\end{abstract}

\section{Introduction}

The light variability of YY CMi ($\equiv$HD67100) was discovered by Morgenroth
(1934). Later photometric observations were due to Lause (1938), Soloviev
(1940), Kaho (1950), and Kordylewsky and Szafraniec (1957). According to the 
General Catalogue of Variable Stars (GCVS, III ed., Kukarkin et al.~1969), 
the system is  classified as a $\beta$ Lyrae type eclipsing binary with a
period of 1.0940253 d and a spectral type F5. Later, the spectrum was classed by 
Hill et al.~(1975) as F6V at phase 0.31 and F7V at phase 0.71. The first
complete light curve in three colours ($u, b, y$) was obtained by Abhyankar
(1962), who also presented a solution based on Russell and Merrill method.
From a questionable treatment of the colour indices, Abhyankar (1962)
concluded that the system is composed of an F6III primary and an A5V secondary.
Koch et al.~(1970) noticed that YY CMi was probably a
system of two (F5 + F8) main sequence stars.

Giuricin \& Mardirossian (1981) reanalyzed Abhyankar's (1962) three--colour
photoelectric observations by using Wood's (1972) model and found
a solution appreciably different from the previous ones. The elements they 
derived lead to an evolved contact system consisting of a primary (roughly an
F6 star) and a secondary (early G5) of practically equal sizes. This picture
of the system is only an approximation of the real one, since Wood's model
treats the stars as triaxial ellipsoids and does not handle contact systems
very well.

\section{Observations}

YY CMi was observed in the framework of a two-site campaign (European
Southern Observatory, La Silla, Chile and Merate Observatory, Italy)
devoted to the $\delta$ Scuti star BI CMi (Mantegazza \& Poretti 1994). The
277 ESO observations cover 14 consecutive nights (from
JD 2448280 to JD 2448293), while the Merate ones are distributed over 8
nights (from JD 2448273 to JD 2448291). We have observations from the two
sites in the same night for 5 cases; since on these nights there is a partial
superimposition of the observations for almost two hours,
it was possible to get an excellent alignment between the two datasets.
The individual $V$ observations are given in Tab. 1 and the respective
light curve is shown in Fig.~1.

\begin{table*}
\caption{Individual $V$ observations of YY CMi}
\begin{tabular}{cccccccccccc}
\hline
HJD  & $\Delta V$  & $HJD$ & $\Delta V$ & $HJD$  & $\Delta V$ &
HJD  & $\Delta V$  & $HJD$ & $\Delta V$ & $HJD$  & $\Delta V$ \\
2448200+&  &2448200+&  &2448200+& &2448200+& &2448200+&  &2448200+& \\ \hline
73.4463&  2.363&  80.4971&  1.886&  82.8047&  2.064&
85.6123&  1.835&  88.7217&  2.093&  90.5918&  1.738 \\
73.4629&  2.424&  80.5039&  1.907&  82.8145&  2.024&
85.6230&  1.804&  88.7305&  2.156&  90.6143&  1.716 \\
73.4688&  2.430&  80.5078&  1.939&  82.8252&  1.984&
85.6787&  1.702&  88.7412&  2.227&  90.6230&  1.710 \\
73.4736&  2.423&  80.5146&  1.949&  82.8359&  1.939&
85.6875&  1.696&  88.7803&  2.429&  90.6318&  1.704 \\
73.4854&  2.367&  80.5215&  2.001&  82.8486&  1.907&
85.6963&  1.690&  88.7891&  2.416&  90.6396&  1.698 \\
73.4941&  2.308&  80.5283&  2.021&  83.5898&  1.629&
85.7070&  1.674&  88.7988&  2.375&  90.6699&  1.679 \\
73.5049&  2.228&  80.5391&  2.081&  83.5986&  1.641&
85.7275&  1.655&  88.8105&  2.293&  90.6787&  1.681 \\
73.5117&  2.192&  80.5439&  2.101&  83.6074&  1.638&
85.7383&  1.644&  88.8213&  2.224&  90.6885&  1.680 \\
73.5186&  2.148&  80.5508&  2.121&  83.6152&  1.646&
85.7471&  1.646&  88.8320&  2.142&  90.7031&  1.679 \\
73.5293&  2.058&  80.5557&  2.139&  83.6240&  1.642&
85.7754&  1.644&  88.8447&  2.061&  90.7109&  1.680 \\
73.5352&  2.025&  80.5654&  2.163&  83.6328&  1.651&
85.7861&  1.636&  89.5400&  1.697&  90.7188&  1.683 \\
73.5400&  1.993&  80.5723&  2.160&  83.6572&  1.669&
85.7949&  1.636&  89.5488&  1.693&  90.7363&  1.687 \\
73.5479&  1.960&  80.5791&  2.173&  83.6807&  1.678&
85.8066&  1.642&  89.5693&  1.688&  90.7432&  1.692 \\
73.5547&  1.922&  80.5889&  2.157&  83.6895&  1.685&
85.8164&  1.650&  89.5781&  1.682&  90.7520&  1.698 \\
73.5645&  1.872&  80.5938&  2.155&  83.6992&  1.706&
85.8291&  1.652&  89.5869&  1.676&  90.7715&  1.710 \\
73.5742&  1.850&  80.6084&  2.091&  83.7080&  1.714&
85.8408&  1.662&  89.5967&  1.680&  90.7793&  1.716 \\
73.5898&  1.796&  80.6172&  2.071&  83.7305&  1.741&
86.5840&  2.415&  89.6172&  1.677&  90.7891&  1.740 \\
73.6064&  1.780&  80.6250&  2.034&  83.7383&  1.758&
86.5947&  2.438&  89.6348&  1.682&  90.7988&  1.740 \\
74.3906&  1.729&  80.6523&  1.935&  83.7471&  1.784&
86.6035&  2.413&  89.6465&  1.690&  90.8086&  1.749 \\
74.4023&  1.747&  80.6592&  1.914&  83.7559&  1.802&
86.6143&  2.360&  89.6563&  1.696&  90.8193&  1.763 \\
74.4160&  1.776&  80.6660&  1.886&  83.7646&  1.827&
86.6221&  2.311&  89.6719&  1.711&  90.8311&  1.793 \\
74.4258&  1.790&  80.6953&  1.808&  83.7900&  1.929&
86.6309&  2.254&  89.6797&  1.717&  90.8447&  1.824 \\
74.4326&  1.819&  80.7051&  1.789&  83.7998&  1.969&
86.6465&  2.134&  89.6885&  1.725&  90.8584&  1.873 \\
74.4473&  1.849&  80.7129&  1.773&  83.8105&  2.019&
86.6592&  2.040&  89.6963&  1.736&  91.3711&  1.692 \\
74.4561&  1.885&  80.7363&  1.749&  83.8213&  2.072&
86.6826&  1.923&  89.7178&  1.762&  91.3838&  1.728 \\
74.4658&  1.928&  80.7441&  1.735&  83.8330&  2.104&
86.6934&  1.879&  89.7256&  1.772&  91.3916&  1.736 \\
74.4727&  1.950&  80.7549&  1.730&  83.8447&  2.157&
86.7041&  1.841&  89.7432&  1.798&  91.4258&  1.829 \\
74.4805&  2.001&  80.7656&  1.719&  84.3760&  2.293&
86.7139&  1.813&  89.7656&  1.867&  91.4287&  1.848 \\
74.4941&  2.072&  80.7793&  1.713&  84.3838&  2.385&
86.7344&  1.760&  89.7744&  1.899&  91.5371&  2.139 \\
74.5059&  2.140&  80.7891&  1.690&  84.4023&  2.425&
86.7490&  1.740&  89.7822&  1.918&  91.5469&  2.111 \\
74.5107&  2.167&  80.8135&  1.686&  84.4150&  2.420&
86.7822&  1.699&  89.7910&  1.944&  91.5576&  2.067 \\
74.5215&  2.258&  80.8271&  1.681&  84.4199&  2.390&
86.7969&  1.675&  89.8008&  2.006&  91.5664&  2.032 \\
74.5273&  2.311&  80.8408&  1.675&  84.4287&  2.349&
86.8066&  1.667&  89.8105&  2.062&  91.5762&  1.988 \\
74.5371&  2.369&  81.5938&  1.890&  84.4375&  2.293&
86.8184&  1.658&  89.8223&  2.136&  91.5947&  1.916 \\
74.5488&  2.411&  81.6025&  1.928&  84.4473&  2.217&
86.8301&  1.654&  89.8369&  2.237&  91.6094&  1.876 \\
74.5547&  2.440&  81.6123&  1.974&  84.4541&  2.157&
86.8438&  1.649&  89.8496&  2.329&  91.7744&  1.677 \\
74.5625&  2.465&  81.6211&  2.012&  84.4629&  2.108&
87.5811&  1.868&  89.8604&  2.369&  91.7842&  1.681 \\
74.5674&  2.438&  81.6289&  2.050&  84.4736&  2.041&
87.5908&  1.912&  90.3613&  1.940&  91.7939&  1.677 \\
74.5742&  2.426&  81.6377&  2.085&  84.4814&  1.985&
87.6006&  1.951&  90.3691&  1.968&  91.8047&  1.678 \\
74.5879&  2.329&  81.6475&  2.116&  84.4883&  1.949&
87.6104&  1.987&  90.3740&  2.000&  91.8174&  1.684 \\
74.5938&  2.297&  81.6563&  2.148&  84.4932&  1.934&
87.6182&  2.035&  90.3848&  2.036&  91.8301&  1.688 \\
74.5977&  2.254&  81.7021&  2.098&  84.5010&  1.896&
87.6289&  2.096&  90.3926&  2.075&  91.8438&  1.691 \\
74.6035&  2.214&  81.7148&  2.046&  84.5166&  1.842&
87.6445&  2.216&  90.4004&  2.102&  91.8574&  1.700 \\
77.3975&  1.846&  81.7344&  1.973&  84.5186&  1.839&
87.6533&  2.266&  90.4053&  2.112&  92.5342&  1.886 \\
77.4063&  1.823&  81.7422&  1.944&  84.5264&  1.806&
87.6621&  2.329&  90.4121&  2.144&  92.5625&  2.018 \\
77.4121&  1.809&  81.7510&  1.913&  84.5371&  1.807&
87.6826&  2.421&  90.4141&  2.151&  92.5723&  2.061 \\
77.4297&  1.767&  81.7607&  1.878&  84.5459&  1.758&
87.6934&  2.424&  90.4209&  2.150&  92.5811&  2.089 \\
77.4414&  1.756&  81.7695&  1.852&  84.5850&  1.706&
87.7041&  2.378&  90.4277&  2.149&  92.6055&  2.167 \\
77.4492&  1.743&  81.7988&  1.791&  84.5938&  1.697&
87.7275&  2.222&  90.4336&  2.145&  92.6133&  2.172 \\
77.4590&  1.730&  82.3770&  1.717&  84.6035&  1.684&
87.7383&  2.158&  90.4414&  2.127&  92.6201&  2.165 \\
77.4736&  1.719&  82.3867&  1.706&  84.6133&  1.672&
87.7490&  2.077&  90.4453&  2.120&  92.6328&  2.139 \\
77.4893&  1.704&  82.3877&  1.710&  84.6211&  1.664&
87.7744&  1.932&  90.4512&  2.115&  92.6436&  2.097 \\
77.4971&  1.709&  82.4736&  1.638&  84.6309&  1.657&
87.7852&  1.889&  90.4580&  2.080&  92.6631&  2.024 \\
77.5039&  1.702&  82.4873&  1.629&  84.6396&  1.653&
87.7979&  1.845&  90.4639&  2.047&  92.6729&  1.987 \\
77.5156&  1.689&  82.5000&  1.628&  84.6611&  1.642&
87.8105&  1.802&  90.4707&  2.024&  92.6924&  1.906 \\
77.5273&  1.691&  82.5244&  1.642&  84.6826&  1.640&
87.8232&  1.774&  90.4746&  2.009&  92.7002&  1.887 \\
77.5332&  1.690&  82.5381&  1.644&  84.6904&  1.633&
87.8369&  1.761&  90.4795&  1.998&  92.7100&  1.858 \\
77.5371&  1.684&  82.5527&  1.667&  84.6992&  1.638&
87.8516&  1.727&  90.4834&  1.970&  92.7314&  1.800 \\
\hline
\end{tabular}
\end{table*}

\begin{table*}
{\bf Table 1} (contd.)\\
\begin{tabular}{cccccccccccc}
\hline
HJD  & $\Delta V$  & $HJD$ & $\Delta V$ & $HJD$  & $\Delta V$ &
HJD  & $\Delta V$  & $HJD$ & $\Delta V$ & $HJD$  & $\Delta V$ \\
2448200+&  &2448200+&  &2448200+& &2448200+& &2448200+&  &2448200+& \\ \hline
77.5498&  1.679&  82.6299&  1.733&  84.7080&  1.641&
88.5439&  1.690&  90.4902&  1.949&  92.7803&  1.730 \\
77.5605&  1.681&  82.6396&  1.751&  84.7334&  1.656&
88.5547&  1.688&  90.4961&  1.934&  92.7900&  1.732 \\
77.5693&  1.665&  82.6484&  1.768&  84.7422&  1.657&
88.5635&  1.699&  90.5049&  1.905&  92.8008&  1.721 \\
77.5859&  1.690&  82.6650&  1.818&  84.7520&  1.663&
88.5859&  1.715&  90.5088&  1.883&  92.8154&  1.708 \\
77.5967&  1.686&  82.6738&  1.838&  84.7646&  1.679&
88.5967&  1.733&  90.5166&  1.855&  92.8291&  1.695 \\
80.3955&  1.669&  82.6934&  1.917&  84.7891&  1.696&
88.6074&  1.739&  90.5205&  1.852&  92.8438&  1.685 \\
80.4131&  1.705&  82.7021&  1.955&  84.7979&  1.706&
88.6162&  1.755&  90.5264&  1.860&  92.8574&  1.680 \\
80.4268&  1.706&  82.7109&  1.996&  84.8076&  1.718&
88.6367&  1.778&  90.5303&  1.851&  93.5361&  1.691 \\
80.4453&  1.730&  82.7207&  2.031&  84.8174&  1.733 &
88.6572&  1.826&  90.5352&  1.828&  93.5488&  1.706  \\
80.4502&  1.746&  82.7412&  2.120&  84.8262&  1.747&
88.6660&  1.845&  90.5420&  1.811&  93.5576&  1.713 \\
80.4600&  1.748&  82.7500&  2.154&  84.8389&  1.771&
88.6846&  1.906&  90.5479&  1.796&  93.5664&  1.727 \\
80.4688&  1.790&  82.7588&  2.169&  85.5850&  1.937&
88.6934&  1.944&  90.5508&  1.792&           &        \\
80.4824&  1.833&  82.7676&  2.168&  85.5938&  1.898&
88.7031&  1.984&  90.5586&  1.784&           &        \\
80.4902&  1.857&  82.7949&  2.103&  85.6035&  1.870&
88.7129&  2.041&  90.5840&  1.742&           &        \\
\hline
\end{tabular}
\end{table*}

We have performed differential photometry in the $V$ band with respect to the
two comparison stars HD 66925 and HD 67028; since the light variability
of the $\delta$ Sct star BI CMi was faster than that of YY CMi, the latter
was measured once every five cycles. The comparison of the differential
magnitudes between the two comparison stars has shown, as expected, a
different accuracy between the data gathered at La Silla and at Merate:
the former have a mean standard deviation of 4.4 mmag for each measurement,
against the 8.6 mmag for the latter (a value quite high due to unfavourable
declination of the field with respect to the latitude of Merate Observatory).
Moreover, Mantegazza \& Poretti (1994) discussed the possible
microvariability of HD 67028.
\section{The period of the system from the times of minima}

\begin{figure}
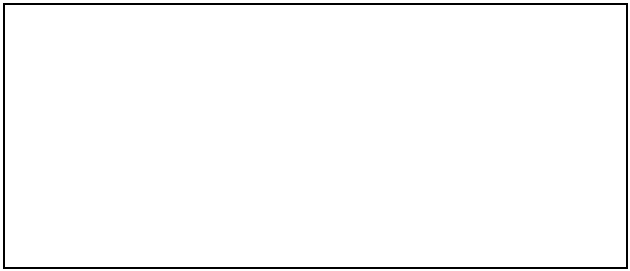

\picplace{3.5cm}
\caption[]{The individual $V$ observations of YY CMi.}
\end{figure}

The present observations, obtained at Merate Observatory and ESO, were 
used to calculate new times of minima by using Kwee and Van Woerden (KW)
method. The new minima times are given in Tab. 2. The successive columns
give the HJD of minimum, the error, the number of cycles $E$, the
$(O - C)$ values and the type of minimum (I: primary, II: secondary). 
A least--squares solution, applied to all minima listed in Tab. 2, yields
the following ephemeris
\begin{eqnarray*}
{\rm J.D.Hel.~(Min \, I)} & = & 2448287.6836 + 1.0937869\cdot E \\
                          &   & \phantom{000\,\,}\pm 0.0023 \pm 0.0003484
\end{eqnarray*}
which has been used for the calculation of $O - C$ values.

The $O-C$ behaviour for all the existing minima times, computed by using the 
above ephemeris, is shown in Fig.~2. The existing data are not enough to 
draw definite conclusions about the variation of the period. The GCSV, IV ed.,
(Kholopov et al.~1985) gives a period 1.0940197 d. This period was
calculated by Abhyankar (1962) using well determined minima times over
a period of 24 years. Our new ephemeris, based on the present observations,
suggests a shorter period.
\begin{table}
\caption{New times of minima of YY CMi}
\begin{tabular}{llcrc}
\hline
  JD Hel.      &    Error   & $E$ &   $O-C$    & Type of minimum \\
  24 40000     &            &     &            &                 \\
\hline
\\

8273.4650      &   0.0050   & -13.0 &  0.0007   &   I \\
8274.5561      &   0.0006   & -12.0 & -0.0020   &   I \\
8280.5688      &   0.0082   & -6.5  & -0.0052   &  II \\
8281.6774      &   0.0044   & -5.5  & 0.0097    &  II \\
8282.7544      &   0.0041   & -4.5  & -0.0071   &  II \\
8284.4080      &   0.0050   & -3.0  & 0.0058    &   I\\
8287.6896      &   0.0030   &  0.0  &  0.0060   &   I \\
8288.7719      &   0.0023   &  1.0  & -0.0055   &   I \\
8290.4151      &   0.0018   &  2.5  & -0.0029   &  II \\
8292.6061      &   0.0007   &  4.5  & 0.0005   &  II \\
\hline
\end{tabular}
\end{table}
\begin{figure}
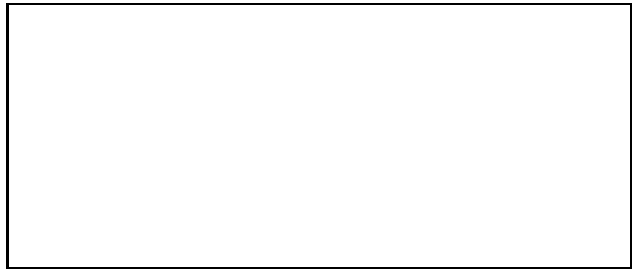

\picplace{3.5cm}
\caption[]{The O - C diagram of YY CMi.}
\end{figure}

\section{Light curve analysis}

The light curve analysis is quite difficult for the following
reasons: (a) no spectroscopic mass-ratio is known; (b) the maxima of the
light curve are unequal in brigthness (Max\, I brighter than Max\, II);
(c) the system undergoes only partial eclipses.
An inspection of the light curve reveals that brightness variations occur
not only around the maxima, but also at other phases. More 
specifically, a decrease in brightness is present in 
the phase interval $0.59-0.87$  and a small excess of
light is seen around phase 0.32. Other minor light 
variations can be seen in other phase regions. The magnitude difference
between the two maxima is about Max\,II - Max\,I = 0.03 mag. 
In modelling light curves of systems exhibiting light curve anomalies,
the need to place hot and/or cool spots of 
solar type has been suggested by several investigators (e.g.~Binnendijk 1960,
Hilditch 1981, Linnell 1982, Van Hamme \& Wilson 1985, Milone et al.~1987,
van't Veer \& Maceroni 1988, 1989, Maceroni et al.~1990).

\begin{table*}
\caption[]{Normal points for YY CMi in light units}
\begin{flushleft}
\begin{tabular}{cccccc}\hline
phase  & $l_{\rm V}$ & $n$ & phase & $l_{\rm V}$ & $n$ \\ \hline
0.00783 &  0.48137 & 11.0 & 0.50773 &  0.61695 &  8.0 \\
0.02287 &  0.51976 & 10.0 & 0.52366 &  0.63518 &  7.0 \\
0.03789 &  0.57334 & 10.0 & 0.53646 &  0.66575 & 10.0 \\
0.05239 &  0.63842 &  8.0 & 0.55349 &  0.71225 & 12.0 \\
0.06562 &  0.69908 &  6.0 & 0.56933 &  0.75813 &  6.0 \\
0.08267 &  0.76389 &  9.0 & 0.58190 &  0.78970 & 11.0 \\
0.09990 &  0.81603 & 10.0 & 0.59689 &  0.82158 &  7.0 \\
0.11422 &  0.85608 &  5.0 & 0.61420 &  0.85730 &  9.0 \\
0.12828 &  0.87877 &  5.0 & 0.62881 &  0.88346 &  4.0 \\
0.14416 &  0.90881 &  3.0 & 0.64715 &  0.90434 &  3.0 \\
0.15985 &  0.93352 &  5.0 & 0.65910 &  0.91634 &  6.0 \\
0.17454 &  0.94462 &  3.0 & 0.67556 &  0.92649 &  5.0 \\
0.18769 &  0.96072 &  5.0 & 0.69124 &  0.93876 &  8.0 \\
0.20482 &  0.97750 &  5.0 & 0.70327 &  0.94778 &  4.0 \\
0.21864 &  0.98643 &  3.0 & 0.71938 &  0.95337 &  6.0 \\
0.23310 &  0.99242 &  4.0 & 0.73380 &  0.96041 &  8.0 \\
0.25381 &  1.00000 &  4.0 & 0.74856 &  0.96248 &  7.0 \\
0.26586 &  0.99982 &  6.0 & 0.76533 &  0.95914 &  5.0 \\
0.27942 &  0.99460 &  5.0 & 0.77979 &  0.95545 &  5.0 \\
0.29477 &  0.98801 &  5.0 & 0.79565 &  0.95080 &  7.0 \\
0.30943 &  0.97813 &  4.0 & 0.80974 &  0.94060 &  4.0 \\
0.32286 &  0.96843 &  3.0 & 0.82615 &  0.92802 &  5.0 \\
0.34133 &  0.96221 &  3.0 & 0.84137 &  0.91143 &  6.0 \\
0.35564 &  0.94087 &  6.0 & 0.85597 &  0.89722 &  4.0 \\
0.37072 &  0.92964 &  7.0 & 0.86969 &  0.87827 &  5.0 \\
0.38562 &  0.90732 &  9.0 & 0.88500 &  0.85045 &  5.0 \\
0.40026 &  0.87768 &  6.0 & 0.90195 &  0.80985 &  6.0 \\
0.41852 &  0.83159 &  7.0 & 0.91745 &  0.76525 &  8.0 \\
0.43452 &  0.78076 &  6.0 & 0.93045 &  0.72516 &  5.0 \\
0.44748 &  0.74103 & 10.0 & 0.94591 &  0.66138 &  8.0 \\
0.46273 &  0.69124 & 11.0 & 0.96274 &  0.58903 &  6.0 \\
0.47808 &  0.64879 & 10.0 & 0.97808 &  0.52715 &  8.0 \\
0.49287 &  0.62114 & 10.0 & 0.99542 &  0.48774 &  7.0 \\
\hline
\end{tabular}
\end{flushleft}
\end{table*}

\subsection{Unspotted solution}

The most recent (1996) version of the Wilson-Devinney (Wilson 1990)
synthetic light curve code was used for the light curve solution. 66 normal 
points, listed in Tab. 3, were used and weights equal
to the number of observations per normal were assigned. Both unspotted and
spotted solutions were performed; for the latter,
we assumed the presence of cool and hot spots to explain the difference in
brightness between the two maxima and the excess of light, respectively.
Under these assumptions we excluded  the observations in the phase interval
$0.59-0.87$ from the unspotted solution, since  a significant
decrease of brightness occurs.

The subscripts 1 and 2 refer to the component eclipsed at primary and
secondary minimum, respectively. A preliminary set of input parameters for
the DC program was obtained by the Binary Maker 2.0 program (Bradstreet 1993).
The DC program was used in the contact mode 3 and in the semidetached mode 4.
In the subsequent analysis
the following assumptions were made: a mean surface temperature 
$T_1=6360~\rm K$ according to the spectral type F6V; we assigned typical values
for stars with convective envelopes to bolometric albedos and gravity darkening
coefficients;
limb darkening coefficients were taken from Al-Naimiy's (1978) tables and 
bolometric linear limb darkening coefficients from Van Hamme (1993). Third
light was assumed to be $\ell_3=0$. The adjustable parameters were: the
phase of conjunction $\phi_0$, the inclination $i$, the temperature $T_2$,
the nondimensional potential $\Omega_1$ in mode 3 and $\Omega_2$ in mode 4,
the monochromatic luminosity $L_1$ and the mass-ratio $q=m_2/m_1$.

Since no spectroscopic mass-ratio of the system is known, a search for the
solution was made for a mass-ratio $q$ ranging from 0.2 to 4. The lowest
values of the sum $\Sigma(\rm res)^2$ of the
weighted squared residuals occured around  $q=1.0$ in mode 3 and
$q=0.8$ in mode 4.
Figure 3 shows the fit parameters $\Sigma(\rm res)^2$ as a function of
the mass-ratio $q$ in modes 3 and 4.
In order to find the final unspotted solution we
continued the analysis by applying the DC program for both cases.
The two solutions converged to $q=0.8921$ in mode 3 and $q=0.8295$ in mode 4. 
The corresponding values of $\Sigma(\rm res)^2$ were found
to be 0.0785 and 0.0835, respectively. Of these two solutions, we finally
adopted the solution in mode 3 (with $q=0.8921$) by taking into account
the better fit of the solution in mode 3 and the fact that the secondary
exceeds the Roche lobe ($\Omega_2 < \Omega_{in}$) in mode 4.
The results of the unspotted solution are given in Table 4 and the
corresponding theoretical light curves are shown as dashed lines in Fig.~4.

\begin{figure}
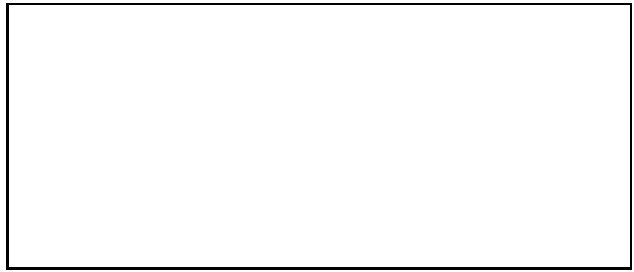

\picplace{3.5cm}
\caption[]{The fit parameter $\Sigma(\rm res)^2$ as a function of the
mass-ratio $q$. Solid lines: mode 3; dashed lines: mode 4.}
\end{figure}

\begin{figure}
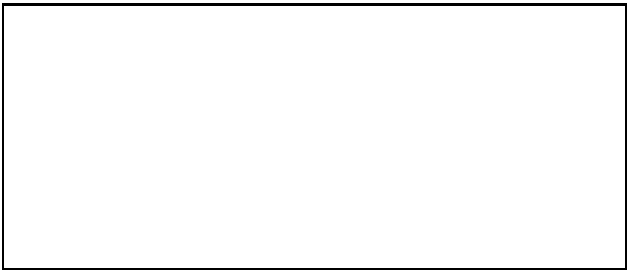

\picplace{3.5cm}
\caption[]{Normal points and theoretical $V$ light curves of YY CMi.
Dashed lines: unspotted solution; solid lines: spotted solution.}
\end{figure}

\begin{table}
\caption{Light curve solutions of YY CMi}
\begin{tabular}{lcc}\hline
Parameter      &   unspotted            &   spotted \\
               &   solution             &   solution \\ \hline  
$\phi_0$       &   $0.0026\pm 0.0003$   &   $0.0026\pm 0.0003$ \\
$i$ (degrees)  &   $79.50\pm 0.16$    &   $79.47\pm 0.07$  \\
$g_1(=g_2)$    &   $0.32^{\ast}$        &   $0.32^{\ast}$  \\
$T_1$ (K)      &   $6360^{\ast}$        &   $6360^{\ast}$  \\
$T_2$ (K)      &   $5707\pm 11$         &   $5710\pm 4$  \\
$A_1$ $(=A_2)$ &   $0.5^{\ast}$         &   $0.5^{\ast}$  \\
$\Omega_1$ $(=\Omega_2)$ & $3.558\pm 0.013$ & $3.560\pm 0.006$ \\
$q=m_2/m_1$    &  $0.892\pm 0.011$  & $0.885\pm 0.004$  \\
$L_1/(L_1+L_2)$ ($V$) & $0.641\pm 0.003$ & $0.642\pm 0.002$ \\
$x_1$ $(=x_2)$  ($V$) & $0.60^{\ast}$    & $0.60^{\ast}$ \\
$x_1$ $(=x_2)$ (bolo) & $0.50^{\ast}$  & $0.50^{\ast}$ \\
\% overcontact &  3\% & 0.3\% \\
$r_1$ (pole)   & $0.368\pm 0.001$     & $0.367\pm 0.001$ \\
$r_1$ (side)   & $0.387\pm 0.001$     & $0.386\pm 0.001$ \\
$r_1$ (back)   & $0.418\pm 0.002$     & $0.416\pm 0.001$ \\
$r_2$ (pole)   & $0.348\pm 0.003$     & $0.346\pm 0.001$ \\
$r_2$ (side)   & $0.366\pm 0.004$     & $0.363\pm 0.002$ \\ 
$r_2$ (back)   & $0.398\pm 0.006$     & $0.395\pm 0.003$ \\ 
$\Sigma_(\rm res)^2$ & 0.0785  &  0.0137 \\ \hline
${\cal M}_1/{\cal M}_{\odot}$ & &  $1.25$ \\
${\cal M}_2/{\cal M}_{\odot}$ & &  $\phantom{^{\ast}}1.12^{\ast}$ \\
$R_1/R_{\odot}$ & & $2.32$ \\
$R_2/R_{\odot}$ & & $2.20$ \\
$\log (L_1/L_{\odot})$ & &  0.91 \\
$\log (L_2/L_{\odot})$ & &  0.67 \\ \hline
\end{tabular}

$^{\ast}$assumed
\end{table}

\subsection{Spotted solution}

The spotted solution was carried out by adopting the simplest spot model
with a physical meaning. We started by assuming that
the system had a cool spot on the secondary (cooler) component of the same
nature as solar magnetic spots (Mullan 1975). Such a spot could explain the
decrease of brightness in the phase interval $0.59 - 0.87$.
Another hot spot was assumed on the secondary component near the neck
region in order to match the light excess around phase 0.32. Such a bright
region can be explained as a result of energy transfer
from the primary to the secondary component (Van Hamme \& Wilson 1985).

The {\it Binary Maker 2.0}
program was used to obtain the best fit by adjusting the spot parameters:
the latitude $b$, the longitude $l$, the angular radius $R$ and
the temperature factor $\rm T.F.$ Once the best fit was obtained, the DC
program was used to derive the final solution. The program allows the
adjustment of spot parameters. The results of the spotted solution
are also given in Tab. 4 and the theoretical light curves are shown as solid
lines in Fig.~4. The $\rm O-C$ differences between the observed and
calculated points for the unspotted and spotted solution for the system
are shown in Fig.~5.

The parameters of the cool spot on the primary component are: latitude
$b=90^{\circ}$ (fixed), longitude $l=271.27^{\circ}\pm 0.91^{\circ}$, angular
radius $R=22.33^{\circ}\pm 4.51^{\circ}$
and temperature factor $\rm T.F.=0.84\pm 0.09$. Those of the hot spot are:
latitude $b=90^{\circ}$ (fixed), longitude $l=21.26^{\circ}\pm 1.95^{\circ}$, 
angular radius $R=9.32^{\circ}\pm 2.69^{\circ}$ and temperature 
factor $\rm T.F.=1.19\pm 0.09$.
A three dimensional picture of the
spotted model at phases 0.25 and 0.75 is shown in Fig.~6, while
the cross-sectional surface outline of the system together with the 
respective critical Roche lobes are given in Fig.~7.

\begin{figure}
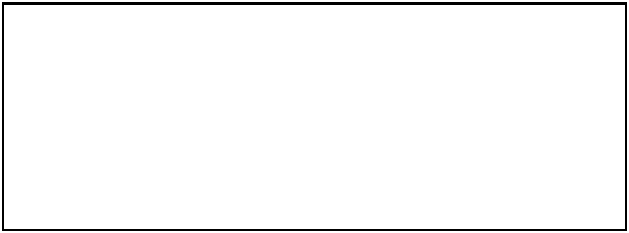
%
\picplace{3.0cm}
\caption[]{The light curve $\rm (O-C)$ residuals for YY CMi in $V$ band.
Crosses refer to unspotted solution; asterisks refer to spotted
solution.}
\end{figure}

\begin{figure}
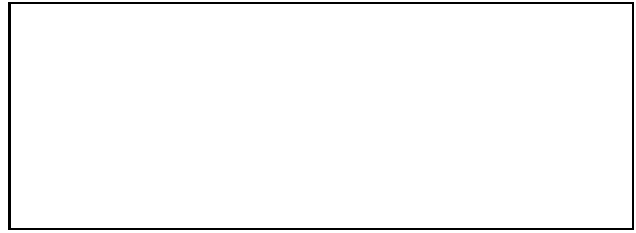

\picplace{3.0cm}
\caption[]{A three-dimensional model of YY CMi for phases 0.25 (upper plot)
and 0.75 (lower plot).}
\end{figure}

\begin{figure}
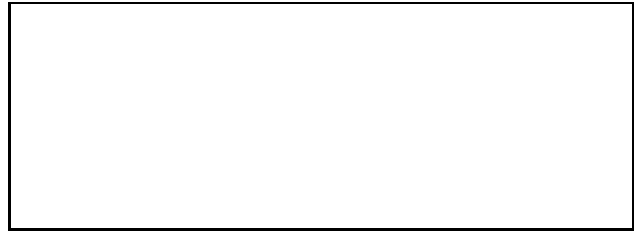

\picplace{3.0cm}
\caption[]{Cross-sectional surface outline of YY CMi. It coincides with the
inner Roche critical surface.}
\end{figure}

\section{Conclusions}

The adopted spot model for YY CMi fits extremely well the observed light 
curves. The absolute elements based on its spectral classification
and the present photometric solution are also given in Tab. 4. We can use
these elements to estimate the evolutionary status by means of the
mass-radius (MR), mass-luminosity (ML) and HR diagrams of Hilditch et
al.~(1988).
In these diagrams, both components of YY CMi lie beyond the
TAMS, in a region occupied mostly by the primaries of A-type W UMa systems.

On the other hand, the degree of contact is almost zero (indicating marginal
contact) and the thermal contact is poor $(T_1~-~T_2)\approx 650 K$. These
results together with the high photometric mass-ratio $q\approx 0.88$ indicate
that YY CMi is very probably a system at
the beginning or the end of the contact phase (Lucy \& Wilson 1979). However,
more definite conclusions about the evolutionary status of YY CMi
can only be drawn by means of new photometric and spectroscopic observations 
of the system. 

\acknowledgements{We thank the referee Dr. R.E. Wilson for his valuable
comments on an earlier version of the manuscript. Figs.~5 and 7 were produced
by \it Binary Maker 2.0.}.

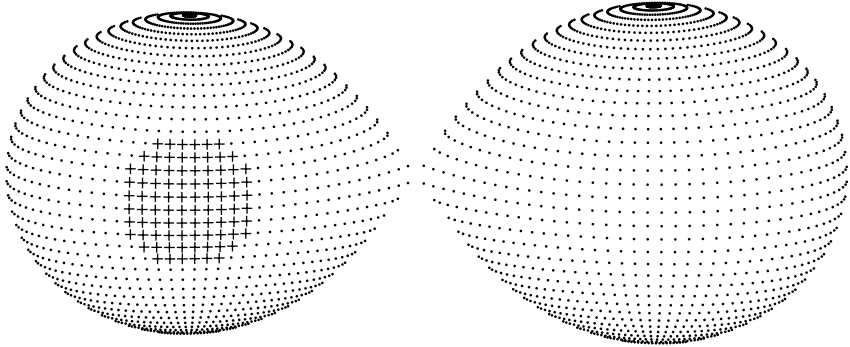
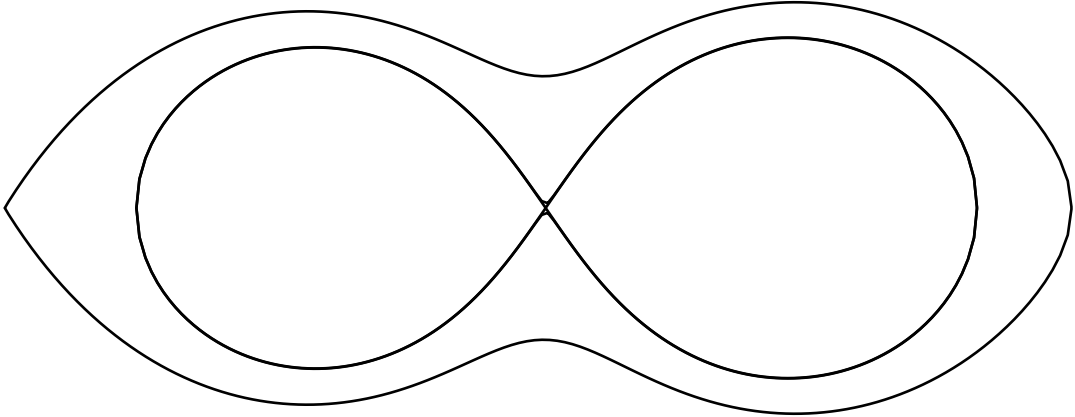
\end{document}